\documentclass[a4paper]{jpconf}
\usepackage{graphicx}
\usepackage{iopams}

\begin{document}
\title{Transition radiation on semi-infinite plate and Smith-Purcell effect}

\author{N F Shul'ga$^1$ and V V Syshchenko$^2$}

\address{$^1$ Akhiezer Institute for Theoretical Physics of the NSC
``KIPT'', Akademicheskaya Street, 1, Kharkov 61108, Ukraine}

\address{$^2$ Belgorod State University, Pobedy Street, 85, Belgorod 308015, Russian Federation}

\ead{shulga@kipt.kharkov.ua, syshch@bsu.edu.ru, syshch@yandex.ru}

\begin{abstract}
The Smith-Purcell radiation is usually measured when an electron
passes over the grating of metallic stripes. However, for high
frequencies (exceeding the plasma frequency of the grating
material) none material could be treated as a conductor, but ought
to be considered as a dielectric with plasma-like permittivity. So
for describing Smith-Purcell radiation in the range of high
frequencies new theoretical approaches are needed. In the present
paper we apply the simple variant of eikonal approximation
developed earlier to the case of radiation on the set of parallel
semi-infinite dielectric plates. The formulae obtained describe
the radiation generated by the particles both passing through the
plates (traditionally referred as ``transition radiation'') and
moving in vacuum over the grating formed by the edges of the
plates (traditionally referred as ``diffraction radiation'', and,
taking into account the periodicity of the plates arrangement, as
Smith-Purcell radiation).
\end{abstract}

\section{Introduction}
An interest to Smith-Purcell effect \cite{SmithPurcell} both as a
prospective way for generation of electromagnetic radiation from
TeraHertz frequencies \cite{THz} to soft X-ray range \cite{Moran}
and as a novel method of beam monitoring (see, e.g.,
\cite{Blackmore}) continuously grows during last years. Usually
the Smith-Purcell radiation is registered when an electron passes
over the grating of metallic stripes. However, for high
frequencies (exceeding the plasma frequency $\omega_p$ of the
grating material) none material could be treated as a perfect
conductor, but ought to be considered as a dielectric with
permittivity $\varepsilon_\omega =1-\omega_p^2/\omega^2$. So for
describing Smith-Purcell radiation in the range of high
frequencies new theoretical approaches are needed.

A simple variant of eikonal approximation suitable for description
of transition radiation on the targets with complex geometry was
proposed in \cite{Syshch13, Syshch14}. In the present paper we
apply that method to the case of radiation on the set of parallel
semi-infinite dielectric plates. The formulae obtained describe
the radiation generated by the particles both passing through the
plates (traditionally referred as ``transition radiation'') and
moving in vacuum over the grating formed by the edges of the
plates (traditionally referred as ``diffraction radiation'', and,
taking into account the periodicity of the plates arrangement, as
Smith-Purcell radiation).

In our article we use the system of units where the speed of light
$c=1$.

\section{Radiation on the semi-infinite plate}
The spectral-angular density of the transition radiation (TR)
could be expressed in the form \cite{Syshch13, Syshch14, Syshch12}
$$
{d\mathcal E\over d\omega d\Omega} = {\omega^2\over(8\pi^2)^2}
\left|\mathbf k\times\mathbf I\right|^2, \eqno(1)
$$
where $\mathbf k$ is the wave vector of the radiated wave,
$$
\mathbf I= \int \left(1-\varepsilon_\omega(\mathbf r) \right)
\mathbf E_\omega(\mathbf r) e^{-i\mathbf k\mathbf r} d^3r
,\eqno(2)
$$
$\mathbf E_\omega$ is the Fourier component by time,
$$
\mathbf E_\omega (\mathbf r) = \int_{-\infty}^\infty \mathbf E
(\mathbf r,t)e^{i\omega t} dt,
$$
of the electric field produced by the moving particle in the
substance of the target with the dielectric permittivity
$\varepsilon_\omega(\mathbf r)$.

If $\left| 1-\varepsilon_\omega(\mathbf r) \right| \ll 1$, the
precise value of the field in the target in (2) could be replaced
in the first approximation by non-disturbed Coulomb field of the
uniformly moving particle in vacuum. Although this approximation
permits to calculate the TR characteristics for the targets with
complex geometry \cite{Syshch12}, its applicability is restricted
by the range of extremely high frequencies,
$$
\omega \gg \gamma\omega_p , \eqno(3)
$$
where $\gamma = \sqrt{1-v^2}$ is the particle's Lorentz factor.
For investigation of TR in the range of more soft photons, some
different approximate method is needed.

A simple variant of eikonal approximation in TR theory was
developed in \cite{Syshch13, Syshch14}. In that approach the
component of $\mathbf I$ perpendicular to the particle's velocity
$\mathbf v$ could be written in the form
$$
\mathbf I^{(eik)}_\perp = i{4e\over v^2\gamma} \int d^2\rho\,
e^{-i\mathbf k_\perp\boldsymbol\rho}\, {\boldsymbol\rho\over\rho}
K_1\left({\omega\rho\over v\gamma}\right)
\left\{\exp\left[-i{\omega\over 2} \int_{-\infty}^\infty
(1-\varepsilon_\omega(\mathbf r))\,dz\right] -1\right\}, \eqno(4)
$$
where $K_n(x)$ is the modified Bessel function of the third kind,
$\boldsymbol\rho$ is the component of $\mathbf r$ perpendicular to
$\mathbf v$. The eikonal approximation in TR theory is valid for
$$
\omega\gg\omega_p \eqno(5)
$$
(compare with the condition(3)), but only for small radiation
angles,
$$
\theta\ll 1/\sqrt{a\omega}, \eqno(6)
$$
where $a$ is the thickness of the target (the last constraint
leads to the possibility of taking into account only the
transverse component of $\mathbf I$).

\begin{figure}
\includegraphics[scale=0.75]{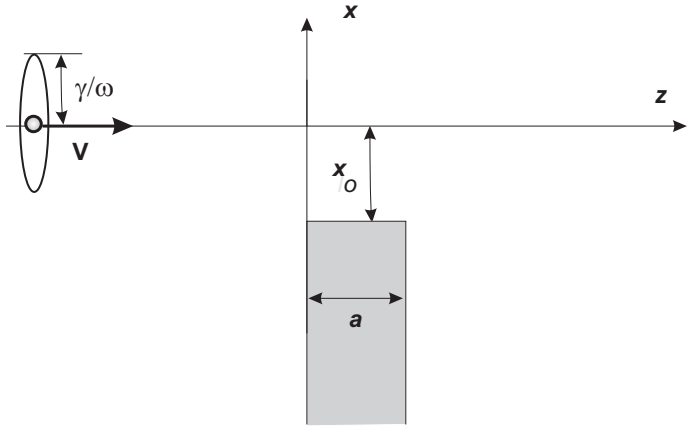} \hspace{1.1pc}
\begin{minipage}[b]{14.5pc}\caption{Particle's motion near the edge of the plate. The value
$\gamma/\omega$ is the characteristic transverse dimension of the
Fourier component of the particle's Coulomb field corresponding to
the frequency $\omega$. Positive values of the impact parameter
$x_0$ correspond to the case of penetration of the particle
through the plate.}
\end{minipage}
\end{figure}

Consider the radiation arising under the interaction of the moving
particle with the semi-infinite dielectric plate of the thickness
$a$ (figure 1). In this case the application of (4) and (1) leads
to the following result for the spectral-angular density of the
radiation:
$$
{d\mathcal E\over d\omega d\Omega} = {e^2\gamma^2\over 2\pi^2}
\left[ 1 - \cos \left( {a\omega\over
2}{\omega_p^2\over\omega^2}\right) \right] F(\theta_x,\theta_y),
\eqno(7)
$$
where $\theta_x$ and $\theta_y$ are the components of the
two-dimensional radiation angle $\boldsymbol\theta$,
$$
F(\theta_x,\theta_y) = {1+2\gamma^2\theta_y^2 \over \left(
1+\gamma^2\theta_y^2 \right)\left( 1+\gamma^2\theta^2 \right)}
\exp\left(-2\left|x_0\right|(\omega/\gamma)\sqrt{1+\gamma^2\theta_y^2}\right)
, \eqno(8a)
$$
under $x_0\leq 0$, and
$$
F(\theta_x,\theta_y) = {4\gamma^2\theta^2 \over
(1+\gamma^2\theta^2)^2} \, + \eqno(8b)
$$
$$
+ {1+2\gamma^2\theta_y^2 \over \left( 1+\gamma^2\theta_y^2
\right)\left( 1+\gamma^2\theta^2 \right)}
\exp\left(-2\left|x_0\right|(\omega/\gamma)\sqrt{1+\gamma^2\theta_y^2}\right)
-
$$
$$
- {4
\exp\left(-\left|x_0\right|(\omega/\gamma)\sqrt{1+\gamma^2\theta_y^2}\right)
\over (+\gamma^2\theta^2)^2} \left( {\gamma\theta_x
\sin(x_0\omega\theta_x) \over \sqrt{1+\gamma^2\theta_y^2}} +
\gamma^2\theta^2 \cos(x_0\omega\theta_x) \right) ,
$$
under $x_0>0$. This formula is in agreement with the result of the
paper \cite{Tishch} obtained using the method \cite{Durand}.

Surface plots of the function $F(\theta_x,\theta_y)$ for some
values of the impact parameter $x_0$ are presented on the figure
2.

\begin{figure}
\includegraphics[scale=0.85]{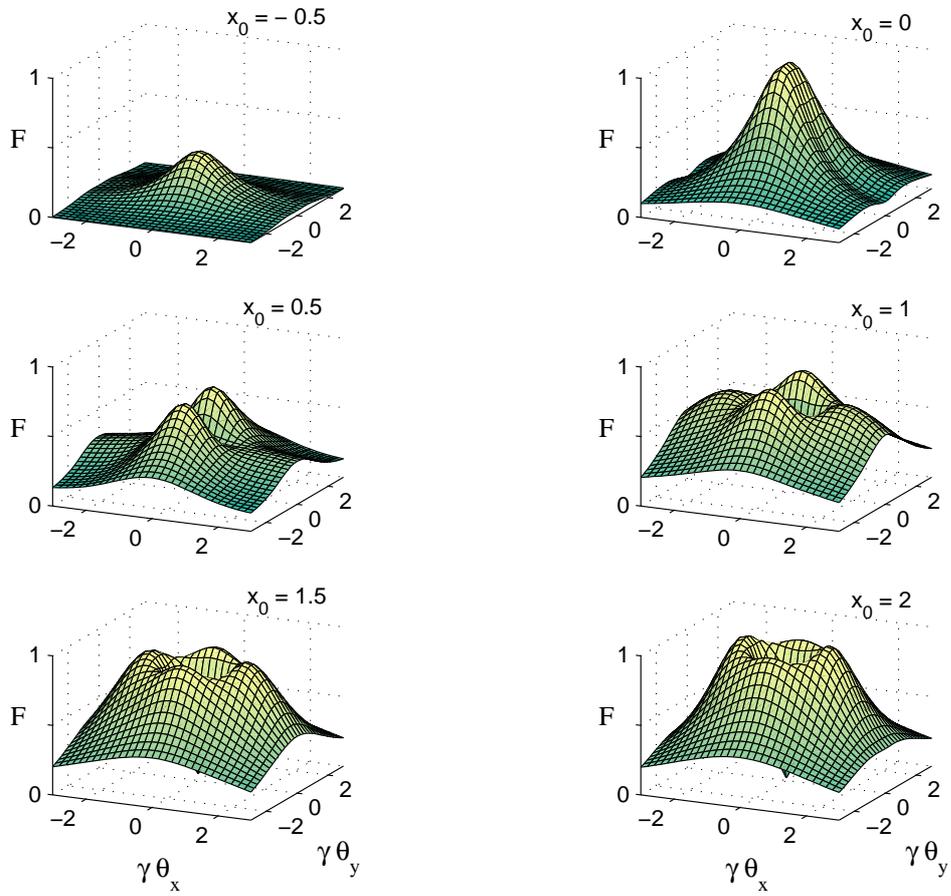}
\caption{Angular distribution of radiation on a single
semi-infinite plate according to (8). Impact parameter $x_0$ is
given in the units of $\gamma/\omega$.}
\end{figure}

\section{Radiation on the grating}
Consider now the radiation arising under incidence on the periodic
set of such plates (figure 3).

\begin{figure}
\includegraphics[scale=0.75]{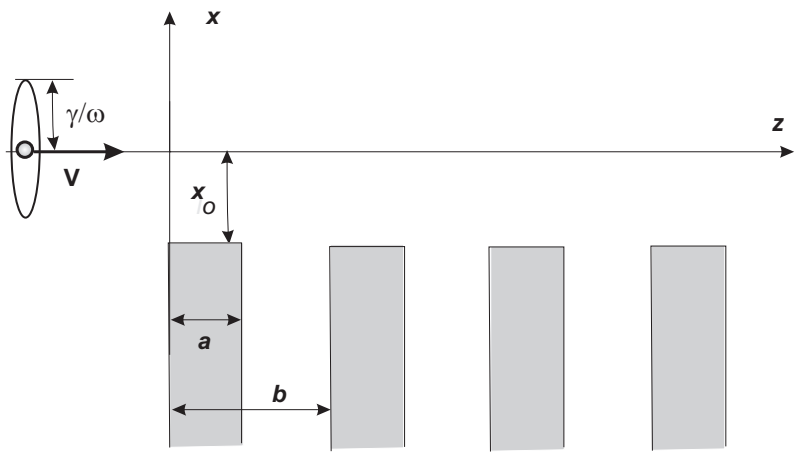} \hspace{1.5pc}
\begin{minipage}[b]{9.5pc}
\caption{Grazing incidence of the particle onto the grating}
\end{minipage}
\end{figure}

Direct account of the periodic structure of the target is
impossible because of the integration along the $z$ axis in the
exponent in (4). Nevertheless, if the formation length of the
radiation on the single plate do not exceed the distance between
two plates,
$$
l_{coh} = {2\gamma^2/\omega \over 1+\gamma^2\theta^2
+\gamma^2\omega_p^2/\omega^2} < b-a , \eqno(9)
$$
the problem is easy for computation: the effect of the target's
complexity will consist in the interference of the electromagnetic
waves emitted under the interaction of the moving particle with
separated plates.

The radiation in this case would be described by equation (7)
multiplied by the factor
$$
2\pi N \sum_{m=-\infty}^\infty \delta \left\{ \omega b \left(
{1\over v}-\cos\theta \right) -2\pi m \right\}, \eqno(10)
$$
where $N$ is the total number of plates, $N\gg 1$ (like in the
most part of the papers devoted to TR, we neglect the refraction
of the emitted radiation in the substance of the target). These
$\delta$-functions mean that the radiation under angle $\theta$
would take the place only for the frequencies satisfying the
condition
$$
\omega = {2\pi m\over {\displaystyle b \left( {1\over
v}-\cos\theta \right)}} \eqno(11a)
$$
($m$ is the positive integer) or, for small angles,
$$
\omega_m = {2\gamma^2\over 1+\gamma^2\theta_m^2}\, {2\pi\over b}
\, m. \eqno(11b)
$$
This is well known Smith-Purcell condition \cite{SmithPurcell}.

\section{Conclusion}
Equations (7), (8a), (10) describe the Smith-Purcell radiation in
the high-frequency limit,
$$
\omega \gg \omega_p .
$$
The exponent in (8a) describes the radiation intensity dependence
on the impact parameter. The characteristic values of the last
one, on which the radiation is substantial, are
$$
\left| x_0 \right|_{eff} = \gamma/2\omega .
$$
Note that such the exponent is present in the formulae describing
Smith-Purcell effect independently on the range of frequencies and
the computation method (see, e.g., \cite{Potyl}; the results of
different models are distinct from each other only by
pre-exponential factors).

In the region $x_0 > 0$ when the particle's trajectory penetrates
the plates, the radiation arising is described by equations (7),
(8b), (10). The first term in (8b) describes the transition
radiation on the infinite plate in the eikonal approximation
\cite{Syshch13, Syshch14}, the second term accounts the edge of
the plate (its contribution evidently coincides with Smith-Purcell
radiation), the third term describes the interference of that two
mechanisms.

In the frames of our approach, for both negative and positive
$x_0$ the angular distribution of the radiation is symmetric in
relation to the $(y,z)$ plane.

\ack This work is supported in part by the internal grant of
Belgorod State University.

\end{document}